# A New Rational Approach to the Square Root of 5 [*]

Shenghui Su [1, 3, 4], Jianhua Zheng [2, 5], and Shuwang Lü [2, 6]

[1] College of Computers, Nanjing Univ. of Aeronautics & Astronautics, Nanjing 211106, PRC
[2] Lab. of Information Security, Chinese Academy of Sciences, Beijing 100039, PRC
[3] School of Computers, National University of Defense Technology, Changsha 410073, PRC
[4] School of Cyberspace Security, Nanjing Univ. of Science & Technology, Nanjing 210094, PRC
[5] Lab. of Digital ID, Peking Knowledge Security Engineering Center, Beijing 100083, PRC
[6] Lab. of Computational Complexity, BFID Corporation, Fuzhou 350207, PRC

**Abstract**: In this paper, the authors construct a new type of sequence which is named an extra-super increasing sequence, and give the definitions of the minimal super increasing sequence $\{a_0, a_1, \cdots, a_n\}$ and minimal extra-super increasing sequence $\{z_0, z_1, \cdots, z_n\}$. Find that there always exists a fit $n$ which makes $(z_n / z_{n-1} - a_n / a_{n-1}) = \Phi$, where $\Phi$ is the golden ratio conjugate with a finite precision in the range of computer expression. Further, derive the formula $\sqrt{5} = 2(z_n / z_{n-1} - a_n / a_{n-1}) + 1$, where $n$ corresponds to the demanded precision. Experiments demonstrate that the approach to $\sqrt{5}$ through a term ratio difference is more smooth and expeditious than through a Taylor power series, and convince the authors that $lim_{n \to \infty}(z_n / z_{n-1} - a_n / a_{n-1}) = \Phi$ holds.

## 1  Introduction

In the world of numbers, there exist some interesting phenomena not easy to observe.

Irrational numbers are those which cannot be expressed each as the ratio of two integers, and thus, the decimal expansion of every irrational number is with an infinite non-repeating decimal. It is also known that an irrational numbers can be expressed as a power series, a non-terminating continued fraction, and many other ways [1][2].

The square root of 5 (namely $\sqrt{5}$) is an important irrational number, and sometimes used for computation of the inradius of a specified pentagon and hypotenuse of a specified right triangle. Customarily, an approximation of $\sqrt{5}$ is approached through the partial sum of a power series. In this paper, on the basis of the minimal super increasing sequence and minimal extra-super increasing sequence, a new method of approaching $\sqrt{5}$ is put forward.

Throughout the paper, unless otherwise specified, the sign $\infty$ denotes the infinity, $\phi$ signifies the golden ratio, $\Phi$ signifies the golden ratio conjugate, $\Sigma$ symbolizes an accumulative sum, $\sqrt{x}$ represents the square root of a positive real number $x$, and $|x|$ means the absolute value of a real number $x$.

## 2  Approach to $\sqrt{5}$ through the Partial Sum of a Power Series

We know that the Taylor series of a function $f(x)$ is an infinite sum of terms which are expressed each in terms of the $n$-th derivative of the function at a certain point $x_0$, namely the function $f(x)$ may be expressed as

$$f(x) = \sum_{n=0}^{\infty} (f^n(x_0) / n!)(x - x_0)^n. \tag{1}$$

If $x_0 = 0$, and there exists the $n$-th derivative of a function at the point 0, then, Formula (1) may be transformed into

[*] This work is supported by MOST with Project 2009AA01Z441.  Corresponding email: idology98@gmail.com.  28 Aug 2021.

$$f(x) = \sum_{n=0}^{\infty}(f^n(0)/n!)x^n. \tag{2}$$

Formula (2) is called the Maclaurin series of $f(x)$, which is the special case of the Taylor series [3].

Further, it is easy to infer the binomial series

$$(1+x)^{\alpha} = 1 + \sum_{n=1}^{\infty}(\alpha(\alpha-1)\cdots(\alpha-n+1)/n!)x^n, \tag{3}$$

which is also a type of power series, and converges for $|x| < 1$ for any real or complex number $\alpha$.

For the special case of $\alpha = 1/2$, we obtain the square root function

$$(1+x)^{1/2} = 1 + (1/2)x - (1/8)x^2 + (1/16)x^3 - (5/128)x^4 + (7/256)x^5 - \cdots$$
$$= \sum_{n=0}^{\infty}(((-1)^{n-1}(2n)!)/(4^n(n!)^2(2n-1)))x^n. \tag{4}$$

According to Formula (4), we can approach $\sqrt{5}$ infinitely [4].

First, $\sqrt{5}$ may be written as $(4+1)^{1/2} = (4(1+1/4))^{1/2} = 2(1+1/4)^{1/2}$, and then

$$\sqrt{5} = 2(\sum_{n=0}^{\infty}(((-1)^{n-1}(2n)!)/(4^n(n!)^2(2n-1)))(1/4)^n). \tag{5}$$

Now, we can compute the approximation of $\sqrt{5}$ according to the $n$-th partial sum of Formula (5).

For $n=0$ to 1, $\sqrt{5} \approx 2(1+(1/2)(1/4)) = 2.25$.

For $n=0$ to 2, $\sqrt{5} \approx 2.25 - 2((1/8)(1/4)^2) = 2.234375$.

For $n=0$ to 3, $\sqrt{5} \approx 2.234375 + 2((1/16)(1/4)^3) = 2.236328125$.

For $n=0$ to 4, $\sqrt{5} \approx 2.236328125 - 2((5/128)(1/4)^4) = 2.23602294921875$.

For $n=0$ to 5, $\sqrt{5} \approx 2.23602294921875 + 2((7/256)(1/4)^5) = 2.23607635498046875$.

For $n=0$ to 6, $\sqrt{5} \approx 2.23607635498046875 - 2((21/1024)(1/4)^6) = 2.236066341400146484375$.

For $n=0$ to 7, $\sqrt{5} \approx 2.236066341400146484375 + 2((33/2048)(1/4)^7) = 2.23606830835342407226 5625$.

For $n=0$ to 8, $\sqrt{5} \approx 2.236068308353424072265625 - 2((429/32768)(1/4)^8)$
$= 2.2360679088160395622 25341796875$.

For $n=0$ to 9, $\sqrt{5} \approx 2.236067908816039562225341796875 + 2((715/65536)(1/4)^9)$
$= 2.236067992052994668483734130859375$.

For $n=0$ to 10, $\sqrt{5} \approx 2.236067992052994668483734130859375 - 2((2431/262144)(1/4)^{10})$
$= 2.23606797436514170840382575988 76953125$.

Obviously, the bigger $n$ is, the more precise the approximation of $\sqrt{5}$ is, which shows wavy approach.

## 3  Definitions of the Minimal Super Increasing Sequence and the Minimal Extra-super Increasing Sequence

A super increasing sequence was firstly proposed in 1978 by R. C. Merkle and M. E. Hellman [5], and used for the design of the MH knapsack cryptosystems [6].

***Definition 1***: For $n+1$ positive integers $a_0, a_1, \cdots,$ and $a_n$ ($n$ to infinity), if $a_i$ ($1 \leq i \leq n$) satisfies

$$a_i > \sum_{j=0}^{i-1} a_j,$$

then this integer progression is called a super increasing sequence, denoted by $\{a_0, a_1, \cdots, a_n\}$, and shortly $\{a_i\}$.

For example, $\{a_0, a_1, \cdots, a_7\} = \{2, 3, 7, 13, 29, 57, 113, 226\}$ is a super increasing sequence.

***Definition 2***: For $n+1$ positive integers $z_0, z_1, \cdots,$ and $z_n$ ($n$ to infinity), if $z_i$ ($1 \leq i \leq n$) satisfies

$$z_i > \sum_{j=0}^{i-1} (i-j)z_j,$$

then this integer progression is called an extra-super increasing sequence, denoted by $\{z_0, z_1, \cdots, z_n\}$, and shortly $\{z_i\}$.

For example, $\{z_0, z_1, \ldots, z_7\} = \{1, 3, 8, 21, 54, 139, 367, 960\}$ is an extra-super increasing sequence.

***Definition 3***: For $n+1$ positive integers $\underline{a}_0, \underline{a}_1, \cdots,$ and $\underline{a}_n$ ($n$ to infinity), if $\underline{a}_0 = 1$, and $\underline{a}_i$ ($1 \leq i \leq n$) satisfies

$$\underline{a}_i = 1 + \sum_{j=0}^{i-1} \underline{a}_j,$$

then this integer progression is called the minimal super increasing sequence, denoted by $\{a_0, a_1, \cdots, a_n\}$, and shortly $\{a_i\}$.

For example, $\{a_0, a_1, \cdots, a_7\} = \{1, 2, 4, 8, 16, 32, 64, 128\}$ is the 7th minimal super increasing sequence.

**Definition 4**: For $n + 1$ positive integers $z_0, z_1, \cdots,$ and $z_n$ ($n$ to infinity), if $z_0 = 1$, and $z_i$ ($1 \leq i \leq n$) satisfies

$$z_i = 1 + \sum_{j=0}^{i-1} (i-j)z_j,$$

then this integer progression is called the minimal extra-super increasing sequence, denoted by $\{z_0, z_1, \cdots, z_n\}$, and shortly $\{z_i\}$.

For example, $\{z_0, z_1, \cdots, z_7\} = \{1, 2, 5, 13, 34, 89, 233, 610\}$ is the 7th minimal extra-super increasing sequence.

## 4  Approach to √5 through the Minimal Extra-super Increasing Sequence

Originally, an extra-super increasing sequence is intended to devise asymmetrical cryptosystems; however, we find accidentally that there exists a fascinating phenomenon in the minimal extra-super increasing sequence.

### 4.1  Term Ratio of the Minimal Extra-super Increasing Sequence

A term ratio of the minimal super increasing sequence or minimal extra-super increasing sequence is a current term to its preceding term.

The term ratio $\mu_i$ is defined as $\mu_i = a_i / a_{i-1}$ (for $i = 1, 2, \cdots, n, \cdots$).

Conforming to Definition 3, it is not difficulty to understand that the term ratio $\mu_i = a_i / a_{i-1} = 2$ is a constant number.

Similarly, the term ratio $\nu_i$ is defined as $\nu_i = z_i / z_{i-1}$ (for $i = 1, 2, \cdots, n, \cdots$).

Conforming to Definition 4, it is not difficulty to calculate

$\{z_0, z_1, \ldots, z_{16}\}$
 $= \{1, 2, 5, 13, 34, 89, 233, 610, 1597, 4181, 10946, 28657, 75025, 196418, 514229, 1346269, 3524578\}$.

Now, compute the term ratios $\nu_1, \nu_2, \cdots, \nu_{16}$.

As $i = 1$, $\nu_1 = z_1 / z_0 = 2 / 1 = 2$.
As $i = 2$, $\nu_2 = z_2 / z_1 = 5 / 2 = 2.5$.
As $i = 3$, $\nu_3 = z_3 / z_2 = 13 / 5 = 2.6$.
As $i = 4$, $\nu_4 = z_4 / z_3 = 34 / 13 \approx 2.6153846$.
As $i = 5$, $\nu_5 = z_5 / z_4 = 89 / 34 \approx 2.617647058$.
As $i = 6$, $\nu_6 = z_6 / z_5 = 233 / 89 \approx 2.61797752808$.
As $i = 7$, $\nu_7 = z_7 / z_6 = 610 / 233 \approx 2.6180257510729$.
As $i = 8$, $\nu_8 = z_8 / z_7 = 1597 / 610 \approx 2.618032786885245$.
As $i = 9$, $\nu_9 = z_9 / z_8 = 4181 / 1597 \approx 2.61803381340012523$.
As $i = 10$, $\nu_{10} = z_{10} / z_9 = 10946 / 4181 \approx 2.6180339631667065295$.
As $i = 11$, $\nu_{11} = z_{11} / z_{10} = 28657 / 10946 \approx 2.618033985017357938973$.
As $i = 12$, $\nu_{12} = z_{12} / z_{11} = 75025 / 28657 \approx 2.6180339882053250514708S$.
As $i = 13$, $\nu_{13} = z_{13} / z_{12} = 196418 / 75025 \approx 2.6180339886704431856047984$.
As $i = 14$, $\nu_{14} = z_{14} / z_{13} = 514229 / 196418 \approx 2.6180339887383030068527324381$.
As $i = 15$, $\nu_{15} = z_{15} / z_{14} = 1346269 / 514229 \approx 2.618033988748203621343798191076$.
As $i = 16$, $\nu_{16} = z_{16} / z_{15} = 3524578 / 1346269 \approx 2.6180339887496481015309718934329$.

In what follows, we evaluate the term ratio difference $v_i - \mu_i$ (for $i = 1, 2, \cdots, 16$).

When $i = 1$, $v_1 - \mu_1 = 0$.
When $i = 2$, $v_2 - \mu_2 = 0.5$.
When $i = 3$, $v_3 - \mu_3 = 0.6$.
When $i = 4$, $v_4 - \mu_4 = 0.6153846$.
When $i = 5$, $v_5 - \mu_5 = 0.617647058$.
When $i = 6$, $v_6 - \mu_6 = 0.61797752808$.
When $i = 7$, $v_7 - \mu_7 = 0.6180257510729$.
When $i = 8$, $v_8 - \mu_8 = 0.618032786885245$.
When $i = 9$, $v_9 - \mu_9 = 0.61803381340012523$.
When $i = 10$, $v_{10} - \mu_{10} = 0.6180339631667065295$.
When $i = 11$, $v_{11} - \mu_{11} = 0.618033985017357938973$.
When $i = 12$, $v_{12} - \mu_{12} = 0.61803398820532505147085$.
When $i = 13$, $v_{13} - \mu_{13} = 0.6180339886704431856047984$.
When $i = 14$, $v_{14} - \mu_{14} = 0.61803398873830300685 2732438$.
When $i = 15$, $v_{15} - \mu_{15} = 0.618033988748203621343 79819107$.
When $i = 16$, $v_{16} - \mu_{16} = 0.6180339887496481015309718934329$.

From the above, we observe that $lim_{n \to \infty}(v_n - \mu_n)$ is trending toward a certain number.

## 4.2 Term Ratio Difference Corresponding to Golden Ratio Conjugate

There exists a relation between the term ratio difference ($z_n/z_{n-1} - a_n/a_{n-1}$) and the golden ratio conjugate with a finite precision.

### 4.2.1 Golden Ratio Conjugate

Two quantities are in the golden ratio if the ratio of the sum of the two quantities to the larger is the same as the ratio of the larger to the smaller [7][8].

Let $\phi (= a / b$ be a ratio) with $a > b > 0$. Then, $a$ and $b$ are said to be in the golden ratio if

$$(a + b) / a = a / b = \phi,$$

namely

$$1 + 1 / \phi = \phi. \tag{6}$$

Formula (6) indicates clearly that $\phi$ is an irrational constant number, and equals approximately to 1.618033988749 [3].

The irrational $\phi$ has many names. Other names frequently used for the golden ratio are the golden section, golden mean, divine proportion, etc [9].

Again let $\Phi (= b / a)$ be a ratio with $a > b > 0$. Then, we can derive from Formula (6) the equation

$$1 + \Phi = 1 / \Phi. \tag{7}$$

Commonly, $\Phi$ is called the golden ratio conjugate [10]. Furthermore, we have

$$\Phi = \phi - 1, \tag{7'}$$

which can easily be inferred from Formula (6) and (7).

Besides, $\phi$ may be represented as a continued fraction [11]

$$\phi = [1; 1, 1, 1, \cdots], \tag{8}$$

and also represented as an infinite series [12]

$$\phi = 13 / 8 + \sum_{n=0}^{\infty}((-1)^{n+1}(2n + 1)!) / (4^{2n+3}n!(n + 2)!). \tag{9}$$

Evidently, an approximation of $\phi$ can be fetched from Formula (8) or (9).

### 4.2.2 Term Ratio Difference ($z_n/z_{n-1} - a_n/a_{n-1}$) Equaling $\Phi$ with a Finite Precision

Resorting to a computer as well as according to Formula (8) and (9), we can acquire

$$\phi \approx 1.618\,033\,988\,749\,894\,848\,204\,586\,834\,365\,638\,117$$

witch has a decimal precision of 36 digits.

Naturally, with the same decimal precision the golden ratio conjugate

$$\Phi \approx 0.618\,033\,988\,749\,894\,848\,204\,586\,834\,365\,638\,117$$

is brought from Formula (7′).

Here, there exists $n = 40$ which makes $(z_{40}/z_{39} - a_{40}/a_{39}) = \Phi$.

Experiments show that for $\Phi$ with any finite precision in the range of computer expression, we can always find a fitly small or large $n$ which makes $(z_n/z_{n-1} - a_n/a_{n-1})$ namely $(v_n - \mu_n)$ be equal to $\Phi$ or $(z_n - z_{n-1})/z_{n-1}$ be equal to $\phi$. These facts imply that $\lim_{n\to\infty}(z_n/z_{n-1} - a_n/a_{n-1})$ trends toward the golden ratio conjugate, or $\lim_{n\to\infty}((z_n - z_{n-1})/z_{n-1})$ trends toward the golden ratio, which will be discussed further in the near future.

### 4.3 New Rational Approach to √5 through a Term Ratio

In terms of Formula (7), there exists

$$\Phi^2 + \Phi - 1 = 0. \tag{7″}$$

Formula (7″) is a quadratic equation, and its significant root is $\Phi = (\sqrt{5} - 1)/2$. Hence, we have

$$\sqrt{5} = 2\Phi + 1. \tag{10}$$

It should be noted that the irrational $\Phi$ is always truncated to a precision length as $\sqrt{5}$ is computed in practical applications.

Due to $(z_n/z_{n-1} - a_n/a_{n-1}) = \Phi$ with a finite precision, furthermore we have

$$\sqrt{5} = 2(z_n/z_{n-1} - a_n/a_{n-1}) + 1, \tag{10′}$$

where $n$ is selected according to a rough demand for the decimal precision of $\sqrt{5}$.

For instance, when the decimal precision demand is 5 (namely $\sqrt{5} = 2.23606$ which is extracted from Section 2), we select $n = 8$. Then

$$\sqrt{5} = 2(z_8/z_7 - a_8/a_7) + 1 = 2(0.6180327868852459) + 1$$
$$= 2.2360655737704918 \approx 2.23606.$$

Again for instance, when the decimal precision demand is 8 (namely $\sqrt{5} = 2.23606797$ which is also extracted from Section 2), we select $n = 11$. Then

$$\sqrt{5} = 2(z_{11}/z_{10} - a_{11}/a_{10}) + 1 = 2(0.618033985017357938973) + 1$$
$$= 2.236067970034715877946 \approx 2.23606797.$$

From experiment results, we see that Formula (10′) is a better approach to $\sqrt{5}$ than Formula (5), and works more smoothly and more expeditiously.

## 5 Conclusion

The paper constructs a new type of sequence which is named an extra-super increasing sequence, and moreover gives the definitions of the minimal super increasing sequence and minimal extra-super increasing sequence.

For $\Phi$ with any finite precision, there always exists a suitable $n$ which makes $v_n - \mu_n = \Phi$, and going ahead, the formula $\sqrt{5} = 2(z_n/z_{n-1} - a_n/a_{n-1}) + 1$ can be derived, which has theoretical and practical values. Experiments illustrate that the approach to $\sqrt{5}$ through a term ratio difference is more smooth and expeditious than through a power series.

For $\Phi$ with an infinite precision, namely the irrational number $\Phi$, in the light of experiment results,

we are inclined to believe that there is $lim_{n \to \infty}(z_n/z_{n-1} - a_n/a_{n-1}) = \Phi$.

Effectively approaching $\sqrt{5}$ is only one application of extra-super increasing sequences. At the next step, we will explore the cryptography meanings of extra-super increasing sequences, and devise new asymmetrical cryptosystems based on extra-super increasing sequences.

## Acknowledgment

The authors would like to thank the Academicians Jiren Cai, Zhongyi Zhou, Zhengyao Wei, Xicheng Lu, Qinmin Wang, Jinpeng Huai, Huaimin Wang, Yaxiang Yuan, Andrew C. Yao, Binxing Fang, and Xiangke Liao for their important advice and help.

The authors also would like to thank the Professors Jie Wang, Zhiying Wang, Ronald L. Rivest, Moti Yung, Dingzhu Du, Hanliang Xu, Yixian Yang, Yupu Hu, Ping Luo, Maozhi Xu, Wenbao Han, Zhiqiu Huang, Zhihui Wei, Lusheng Chen, Bogang Lin, Yiqi Dai, Lequan Min, Dingyi Pei, Mulan Liu, Huanguo Zhang, Qibin Zhai, Hong Zhu, Renji Tao, Quanyuan Wu, and Zhichang Qi for their important suggestions and corrections.